\documentclass[twocolumn,showpacs,superscriptaddress,prb,aps]{revtex4}
\usepackage{amssymb}
\usepackage{amsmath}
\usepackage{graphicx}

\newcommand{\up}{\uparrow}
\newcommand{\dn}{\downarrow}

\begin{document}

\title{Spin Hall effect in spin-valley coupled monolayer transition-metal dichalcogenides}
\author{Wen-Yu Shan}
\affiliation{Department of Physics, Carnegie Mellon University, Pittsburg, Pennsylvania 15213, USA}
\author{Hai-Zhou Lu}
\affiliation{Department of Physics and Centre of Theoretical and Computational Physics, The University of Hong Kong, Pokfulam Road, Hong Kong, China}
\author{Di Xiao}
\affiliation{Department of Physics, Carnegie Mellon University, Pittsburg, Pennsylvania 15213, USA}

\begin{abstract}
We study both the intrinsic and extrinsic spin Hall effect in spin-valley coupled monolayers of transition metal dichalcogenides.  We find that whereas the skew-scattering contribution is suppressed by the large band gap, the side-jump contribution is comparable to the intrinsic one with opposite sign in the presence of scalar and magnetic scattering.  Intervalley scattering tends to suppress the side-jump contribution due to the loss of coherence.  By tuning the ratio of intra- to intervalley scattering, the spin Hall conductivity shows a sign change in hole-doped samples.  Multiband effect in other doping regime is considered, and it is found that the sign change exists in the heavily hole-doped regime, but not in the electron-doped regime.
\end{abstract}

\pacs{72.10.-d, 72.25.Dc, 73.63.-b, 75.70.Tj}

\maketitle

\section{Introduction}

Monolayers of transition-metal dichalcogenides $MX_2$ ($M=$ Mo, W, $X=$ S, Se) have attracted intense recent interest due to their unique optical and electronic properties.~\cite{overview}  These two-dimensional materials can be regarded as semiconductor analog of graphene: their band structure consists of two degenerate but inequivalent valleys located at the corners of the hexagonal Brillouin zone, with a direct band gap in the visible frequency range.~\cite{Splendiani10nanolett,Mak10prl}  It was predicted that,~\cite{Xiao12prl,Cao12nacom} due to the lack of inversion center in the crystal structure, the two valleys can be distinguished by the Berry phase of the Bloch bands,~\cite{Xiao10rmp} which gives rise to the valley Hall effect and valley-dependent optical selection rule.~\cite{Xiao07prl,Yao08prb}  This prediction has motivated several recent experiments, in which the optical generation~\cite{Zeng12nano,Mak12nano,Cao12nacom} and electric control~\cite{Wu13nphs} of valley polarization have been demonstrated.

\begin{figure}[b]
\centering
\includegraphics[width=0.7\columnwidth]{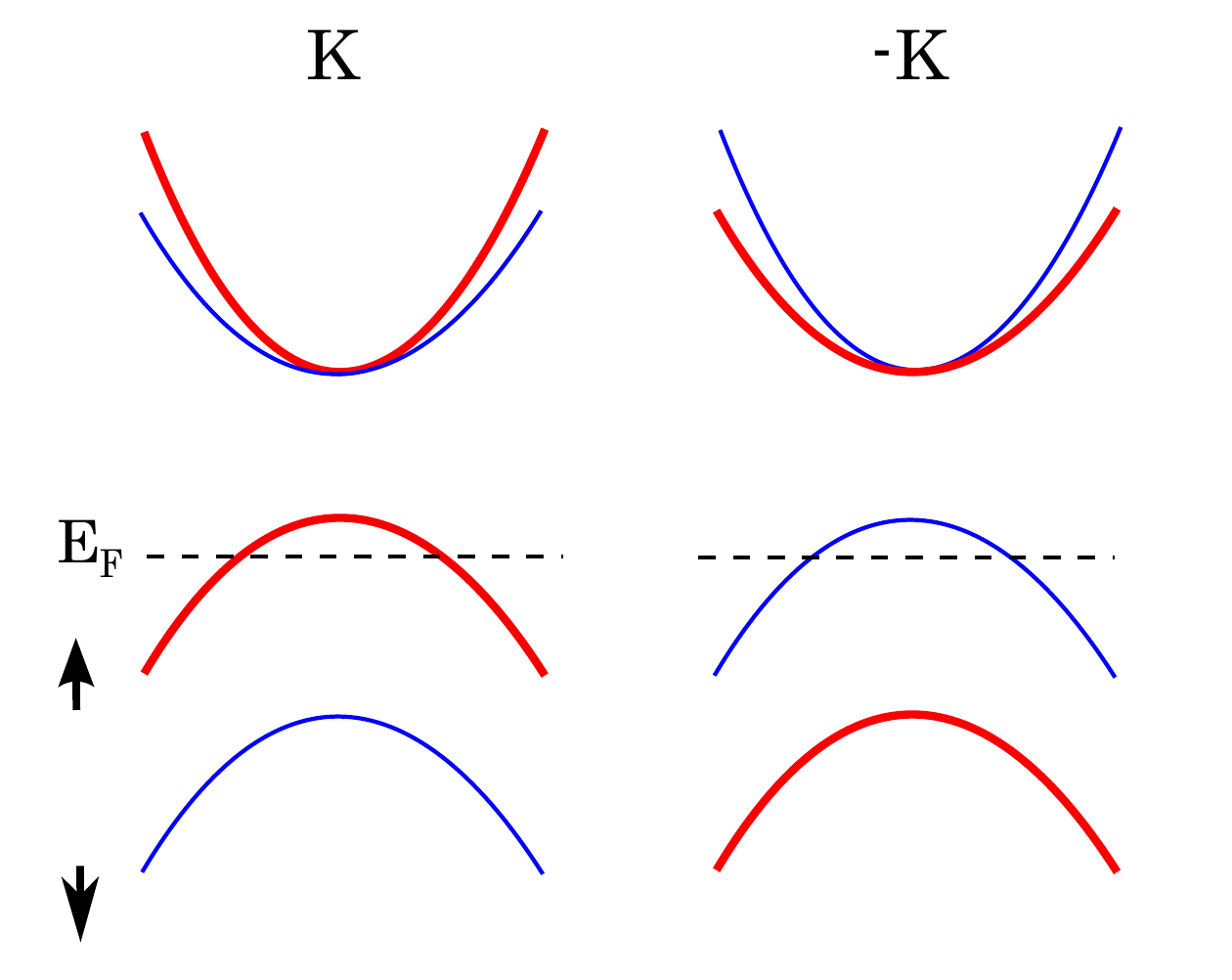}
\caption{(Color online) Schematics of low-energy band structure for monolayer MX$_2$. Red (blue) curves represent bands with spin up (down). Black dashed line shows the Fermi energy $E_F$ measured from the middle of the gap. }\label{fig:band_dispersion}
\end{figure}

Another interesting property of $MX_2$ is the large spin-orbit coupling (SOC) derived from the heavy metal $d$-orbitals.~\cite{Zhu11prb}  It was pointed out that broken inversion symmetry also gives rise to a strong spin-valley coupling,~\cite{Xiao12prl} i.e., carriers in opposite valleys have opposite spin moment (Fig.~\ref{fig:band_dispersion}).  This coupling has a number of implications.  First, various valley-dependent phenomena now become spin-dependent.  In particular, the valley Hall effect is accompanied by a spin Hall effect, in which a transverse spin current can be generated by a longitudinal electric field.  Secondly, the spin-valley coupling dictates that intravalley scattering conserves the spin index whereas intervalley scattering necessarily flips it, resulting in prolonged spin lifetime in the diffusion regime.~\cite{Ochoa13arXiv}  The intra- and inter-valley scattering also leads to opposite localization behavior.~\cite{Lu13prl}

In general, the spin Hall effect consists of both intrinsic and extrinsic contributions.  The intrinsic part, determined by the Berry curvature of the Bloch bands,~\cite{Murakami03science,Sinova04prl} has been discussed for monolayer $MX_2$ based on  first-principles band-structure calculations.~\cite{Feng12prb} On the other hand, it is well known that impurity scattering could modify the intrinsic contribution as demonstrated in the Rashba model.~\cite{Inoue04prb,Raimondi05prb,Rashba04prb} Furthermore, impurity scattering itself may lead to extrinsic spin Hall effect,\cite{Dyakonov71jetplett,Dyakonov71pla,Hirsch99prl} including both skew-scattering~\cite{Smit55physica} and side-jump~\cite{Berger70prb} mechanisms.  Since the strong spin-valley coupling severely limits the possible scattering channels, it is important to investigate its role in the spin Hall effect in $MX_2$ monolayers.

In this work, we calculate the spin Hall conductivity (SHC) of $MX_2$ monolayers. We find that symmetric vertex correction has little effect on the intrinsic contribution due to the large band gap. For the extrinsic mechanisms, the side-jump contribution is comparable to the intrinsic contribution, while the skew scattering contribution is suppressed by the large band gap for both scalar and magnetic scattering. In hole-doped samples (Fig.~\ref{fig:band_dispersion}), the sign of the SHC is opposite for the side-jump and intrinsic contributions, and the side-jump contribution is always suppressed by intervalley scattering due to the loss of coherence. Hence, by tuning the relative strength of intra- and intervalley scattering, the total SHC shows a sign change, i.e., it is negative for weak intervalley scattering and positive for strong intervalley scattering.  Our study is also extended to the multiband case when the system is electron- or heavily hole-doped (such that the Fermi energy crosses multiple bands). We find that the sign change exists in the hole-doped regime, but not in the electron-doped regime.  Therefore, the SHC may provide another measure to determine the strength of intervalley scattering in hole-doped $MX_2$ monolayers.

This paper is organized as follows. In Sec.~\ref{sec:model} we describe the effective model of monolayer $MX_2$ at valley $\pm K$. In Sec.~\ref{sec:disorder_relaxation_vertex} we introduce some important definitions, including scattering potential, relaxation time and vertex correction. The result of the SHC is presented in Sec.~\ref{sec:spin_Hall_hole} and \ref{sec:spin_Hall_electron} for the single-band (lightly hole-doped) and the multi-band case (electron- or heavily hole-doped regime), respectively.  Finally, a discussion and conclusion is given in Sec.~\ref{sec:discussion1}.

\section{\label{sec:model}Model}

Monolayers of $MX_2$ has the crystal symmetry $D_{3h}$. The electronic properties due to the lack of inversion symmetry and large atomic SOC from metal $d$ orbitals are captured by the low-energy effective model around the zone corners $K(-K)$:~\cite{Xiao12prl}
\begin{eqnarray}
H&=&at(\tau_v k_x\hat{\sigma}_x+k_y\hat{\sigma}_y)+\frac{\Delta}{2}\hat{\sigma}_z-\lambda\tau_v\hat{s}_z\otimes\frac{\hat{\sigma}_z-1}{2},
\end{eqnarray}
where $\hat{\sigma}$ and $\hat{s}$ act on the orbital $\{d_{z^2},(d_{x^2-y^2}+i\tau_v d_{xy})/\sqrt{2}\}$ and spin space, respectively. $\otimes$ is the Kronecker product. $\tau_v=\pm 1$ refers to $\pm K$ valley. $\Delta$ is the energy gap, $a$ is the lattice constant, $t$ is the hopping integral, and $\lambda$ is the spin-orbit coupling constant. Note that the complex orbital basis are orthogonal to each other, reducing the coherence of intervalley scattering.

The band dispersion reads
\begin{eqnarray}
E^{m}_{\tau_v,s}&=&\tau_v s\frac{\lambda}{2}\pm m\sqrt{(\frac{\Delta}{2}-\tau_v s\frac{\lambda}{2})^2+a^2t^2k^2},
\end{eqnarray}
where $m,s=\pm1$ correspond to the conduction (valence) band and the spin up (down) state, respectively. The dispersion is shown schematically in Fig.~\ref{fig:band_dispersion}. For each band, eigenfunctions are given by \begin{equation}\label{eigenstates} \begin{split}
|c,\tau_v K,s\rangle &=|s\rangle\otimes
\binom{\chi_n}{
\tau_v w_ne^{i\tau_v\varphi_{\bold{k}}}} \\
|v,\tau_v K,s\rangle &=|s\rangle\otimes
\binom{w_n}{-\tau_v\chi_ne^{i\tau_v\varphi_{\bold{k}}}}
\end{split} \end{equation}
where $c/v$ labels the conduction (valence) band. $n=1,2$ for $\tau_v s=\pm1$, respectively, i.e., $n=1$ for $(K,\uparrow)$ and $(-K,\downarrow)$, and $n=2$ for $(K,\downarrow)$ and $(-K,\uparrow)$. $\chi_{n}$, $w_{n}$ are defined by
\begin{gather}
\chi_{n} = \cos\frac{\theta_{n}}{2},\quad  w_{n}=\sin\frac{\theta_{n}}{2},\\\cos\theta_n=\frac{\frac{\Delta}{2}+(-1)^n\frac{\lambda}{2}}
{\sqrt{(\frac{\Delta}{2}+(-1)^n\frac{\lambda}{2})^2+a^2t^2k^2}},\label{theta_def}
\end{gather}
with $\tan\varphi_{\bold{k}}=k_y/k_x$.

\section{\label{sec:disorder_relaxation_vertex}Disorder, relaxation time and vertex correction}

\subsection{\label{sec:disorder}Impurity potentials}

To calculate the extrinsic SHC, we apply the standard diagrammatic approach, in which the scattering due to impurities and disorders is treated as the perturbation to the eigenstates of $H$.
We consider both scalar and magnetic impurities. Their potentials in real space can be modeled by
\begin{eqnarray}
U(\bold{r})&=&\sum_{i,\alpha=0,x,y,z}u_{\alpha}^i(\sigma_{\alpha}\otimes I)\delta(\bold{r}-\bold{R}_i),
\end{eqnarray}
where $\sigma_{\alpha}$ and $I$ act on the spin and orbital space, respectively. $\bold{R}_i$ and $u$ represent the position and scattering strength of an impurity.
We assume that the impurities have short-range potential and are delta-correlated, i.e., $\langle U(\bold{r})\rangle_{dis}=0$ and $\langle U(\bold{r})U(\bold{r}^{'})\rangle_{dis}=nu^2\delta(\bold{r-r^{'}})$, where $n$ is the disorder concentration.
Although intravalley scattering should be related to long-range potential, the
practice by the delta potential is justified by numerical
calculations.\cite{Yan08prl} In order to include the skew-scattering effect, third-order scattering correlation has to be considered, i.e.,  $\langle U(\bold{r})U(\bold{r}^{'})U(\bold{r}^{''})\rangle_{dis}=nv^3\delta(\bold{r-r^{'}})\delta(\bold{r-r^{''}})$.~\cite{Sinitsyn07prb,Tse06prl}

With the potential and the eigenstates in Eq. (\ref{eigenstates}), the scattering matrix elements for the intravalley scattering are found as
\begin{eqnarray}
U_{\bold{kk}^{'}}&=&\sum_{i,\alpha=0,x,y,z}\frac{u_{\alpha}^i}{S}e^{i(\bold{k^{'}-k})\cdot\bold{R}_i}(\sigma_{\alpha}\otimes I),
\end{eqnarray}
where $S$ is the area of the system. Besides intravalley scattering, we also take into account intervalley scattering induced by short-range disorder. \cite{Suzuura02prl}
The potential for the intervalley scattering is given by
\begin{eqnarray}\label{inter_potential}
&&U^I(\bold{r})\nonumber\\
&=&\sum_{i,\alpha=0,x,y,z}\sigma_{\alpha}\otimes
\left(\begin{array}{cc}
t_{\alpha,A}^i\delta(\bold{r}-\bold{R}_i^{A}) & 0\\
0 & t_{\alpha,B}^i\delta(\bold{r}-\bold{R}_i^{B})\\
\end{array}\right)\nonumber\\
&\otimes&\left(\begin{array}{cc}
0 & e^{-i(\bold{K^{'}-K})\cdot\bold{r}}\\
e^{i(\bold{K^{'}-K})\cdot\bold{r}} & 0\\
\end{array}\right),
\end{eqnarray}
where the basis of matrices represent spin, orbital and valley, respectively, and we have used $A$ and $B$ to label the two orbitals at each valley. For intervalley scattering, we also consider the scalar and magnetic impurities. Note that $B$ orbitals are orthogonal between different valleys, the middle matrix becomes
\begin{eqnarray}
&&\left(\begin{array}{cc}
t_{\alpha,A}^i\delta(\bold{r}-\bold{R}_i^{A}) & 0\\
0 & t_{\alpha,B}^i\delta(\bold{r}-\bold{R}_i^{B})\\
\end{array}\right)\nonumber\\
&\rightarrow&\left(\begin{array}{cc}
t_{\alpha,A}^i\delta(\bold{r}-\bold{R}_i^{A}) & 0\\
0 & 0\\
\end{array}\right).
\end{eqnarray}
When dealing with the intervalley scattering, the first two matrices in Eq. (\ref{inter_potential}) gives the scattering matrix element
\begin{eqnarray}\label{intervalley}
U^I_{\bold{k},\bold{k}^{'}}&=&\sum_{i}
\left(\begin{array}{cccc}
t_0^i+t_z^i & 0 & t_x^i-it_y^i & 0\\
0 & 0 & 0 & 0\\
t_x^i+it_y^i & 0 & t_0^i-t_z^i & 0\\
0 & 0 & 0 & 0\\
\end{array}\right)\frac{e^{i(\bold{k^{'}-k})\cdot\bold{R}_i^A}}{S}.\nonumber\\
\end{eqnarray}
With these scattering matrix elements, the correlation between them can be derived.

\subsection{\label{sec:vertex}Relaxation times}

The scattering will reduce lifetime of the eigenstates of $H$ to finite. The lifetime can be defined with the help of relaxation times. In the single-band case (lightly hole doped), the total relaxation time under the first-order Born approximation reads
\begin{eqnarray}\label{tau}
\frac{1}{\tau}&=&\frac{1}{\tau_{intra}}+\frac{1}{\tau_{inter}},
\end{eqnarray}
with the intravalley $\tau_{intra}$ and intervalley $\tau_{inter}$ defined as
\begin{eqnarray}
\frac{1}{\tau_{intra}}&=&\frac{2\pi}{\hbar}N_1(n_0u_0^2+n_zu_z^2)(w_1^4+\chi_1^4),\\
\frac{1}{\tau_{inter}}&=&\frac{2\pi}{\hbar}N_1(n_xt_x^2+n_yt_y^2)w_1^4,
\end{eqnarray}
where $N_{n}=|\lambda/2+(-1)^{n}E_F|/2\pi a^2t^2$ is the density of states at the Fermi energy $E_F$. Here $n=1$ refers to the highest valence bands located at the two valleys, $(K,\uparrow)$ and $(-K,\downarrow)$. $n_0$ and $n_{x,y,z}$ are the disorder concentration for scalar and magnetic scattering, respectively ($n_{z}$ can be different from $n_{x}$, $n_{y}$ due to the different scattering types).

To include the multi-band effect, we further introduce a set of effective relaxation times $\tau_{(n,p,q)}$. We present the descriptions of the relaxation times in Table \ref{tab:relaxation_times}, and the exact expressions in Appendix \ref{sec:relaxation1}.
It is convenient to use these effective relaxation times to define the relaxation times of physical meanings. For example,
\begin{eqnarray}
\frac{1}{\tau_{intra}}&=&\frac{w_1^4+\chi_1^4}{\tau_{(1,1,1)}},\quad
\frac{1}{\tau_{inter}}=\frac{w_1^4}{\tau_{(1,2,2)}}.
\end{eqnarray}
Later we will see that $\tau_{(1,1,1)}/\tau_{(1,2,2)}$ measures the energy independent ratio between inter- and intravalley scattering.  This parameter will be used throughout the following discussion.

\begin{table}[htbp]
\caption{The descriptions of the effective relaxation times, based on their scattering processes and origins. $0,x,y,z$ indicate the impurity potential that give the relaxation times. $0$ for the scalar scattering, $x,y,z$ are for the three components of the magnetic scattering.}
\label{tab:relaxation_times}%
\begin{ruledtabular}
\begin{tabular}{ccc}
Description &  Spin up (down)  & Spin down (up)  \\
 &  at $K (-K)$  & at $K (-K)$  \\ \hline
Intravalley ($0,z$) & $\tau_{(1,1,1)}$ & $\tau_{(2,1,1)}$  \\
Intravalley ($x,y$) & $\tau_{(1,2,1)}$ & $\tau_{(2,2,1)}$  \\
Intervalley ($0,z$) & $\tau_{(1,1,2)}$ & $\tau_{(2,1,2)}$  \\
Intervalley ($x,y$) & $\tau_{(1,2,2)}$ & $\tau_{(2,2,2)}$  \\
Skew scattering & $\tau_{(1,1,3)}$ & $\tau_{(2,1,3)}$
\end{tabular}
\end{ruledtabular}
\end{table}

\subsection{Vertex correction to Velocity}

One of the direct and important disorder effects on the SHC is due to the vertex correction to velocity.~\cite{Shon98jpsj} A well-known example is for the Rashba model where the vertex correction cancels exactly the intrinsic SHC.~\cite{Inoue04prb,Raimondi05prb,Rashba04prb} In contrast, the spin-valley coupled model with a large band gap considered here gives qualitatively different vertex correction to velocity. This is also different from the discussion on the single-flavor massive Dirac fermions;~\cite{Sinitsyn07prb} here the extra valley degree of freedom and intervalley scattering also modify the vertex correction.

The diagram for the corrected velocity vertex $\widetilde{\mathbf{v}}_y$ is shown in Fig. \ref{fig:diagram} (a). Since the low-energy effective model requires that $k\ll |\bold{K}-(-\bold{K})|$, the valley index should be conserved.\cite{Suzuura02prl} This means only diagonal terms $\widetilde{v}_y^{K}$,$\widetilde{v}_y^{-K}$ (short for $\widetilde{v}_y^{KK}$,$\widetilde{v}_y^{-K-K}$) are possible. Depending on the doping level, one or multiple bands can cross the Fermi level and give different forms of vertex correction.

\begin{figure}[htbp]
\centering
\includegraphics[width=0.8\columnwidth]{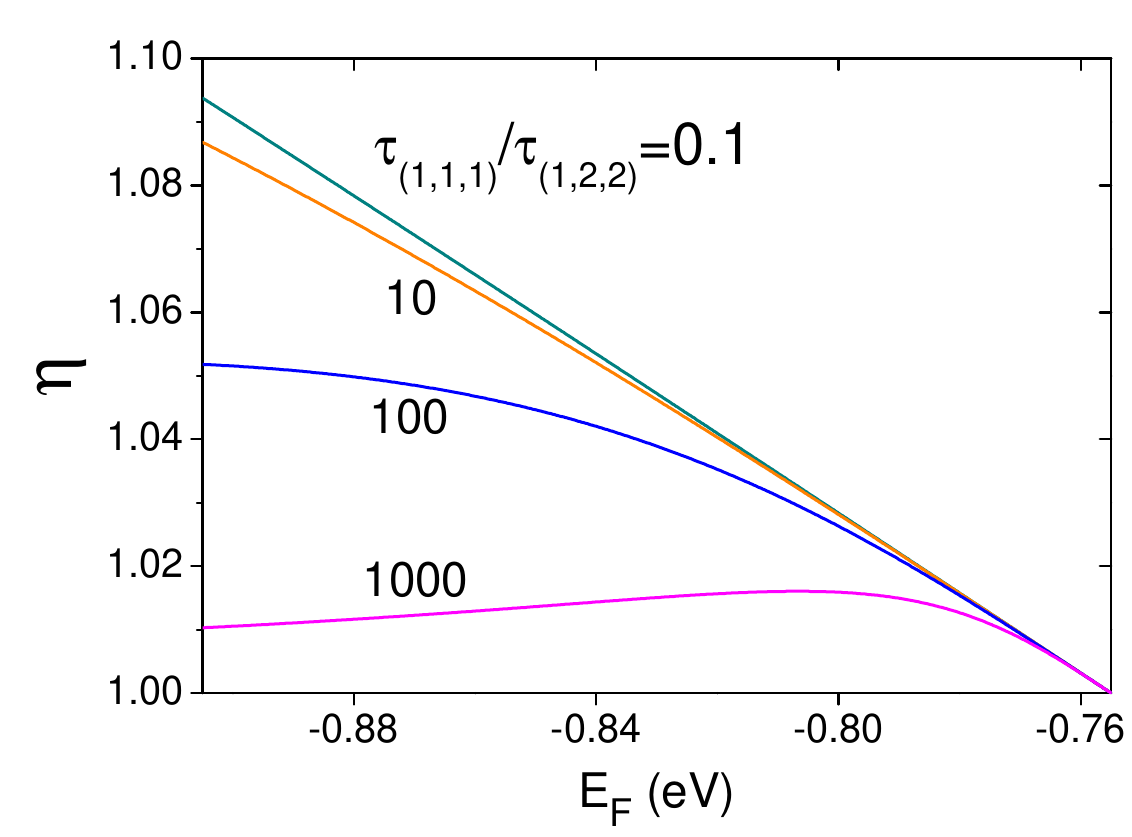}
\caption{The factor $\eta$ that corrects the velocity in the lightly hole-doped regime as functions of Fermi energy $E_F$ for different $\tau_{(1,1,1)}/\tau_{(1,2,2)}$, the ratio of intravalley scattering time to intervalley scattering time. All parameters are adopted for MoS$_2$ from Ref. [\onlinecite{Xiao12prl}].  }\label{fig:vertex1}
\end{figure}
\emph{Single-band case}: In this case, the Fermi level is located in spin-orbit split gap at the valence band top, as shown in Fig. \ref{fig:band_dispersion}.  According to the ladder diagram expansion in Fig. \ref{fig:diagram} (a), a self-consistent equation can be constructed
\begin{eqnarray}
\left(\begin{array}{ccc}
   \widetilde{v}^{K}_{y\bold{k}}\\
   \widetilde{v}^{-K}_{y\bold{k}} \\
    \end{array}\right)&=&\left(\begin{array}{ccc}
   v_{y\bold{k}}^{K}\\
   v_{y\bold{k}}^{-K} \\
    \end{array}\right)+\sum_{\bold{k}^{'}}
\left(\begin{array}{cc}
f(K,K) & f(K,-K)  \\
f(-K,K) & f(-K,-K) \\
    \end{array}\right)\nonumber\\
&\times&\left(\begin{array}{ccc}
   \widetilde{v}^{K}_{y\bold{k}^{'}}\\
   \widetilde{v}^{-K}_{y\bold{k}^{'}} \\
    \end{array}\right),\label{ladder_diagram}
\end{eqnarray}
with the kernel function $f$ defined as
\begin{eqnarray}\label{kernel}
f(\alpha,\beta)&=&
\langle U^{\alpha\beta}_{\bold{k}\bold{k}^{'}}U^{\beta\alpha}_{\bold{k}^{'}\bold{k}}\rangle_{dis}
G^{R}_{\bold{k}^{'},\beta}G^{A}_{\bold{k}^{'},\beta}.
\end{eqnarray}
Due to the particular form of the intervalley scattering in Eq. (\ref{intervalley}), the correlation $\langle U^{\alpha\beta}_{\bold{k}\bold{k}^{'}}U^{\beta\alpha}_{\bold{k}^{'}\bold{k}}\rangle_{dis}$  as well as $f(\alpha,\beta)$ for $\alpha\neq\beta$ become angle independent, implying that the matrix in Eq. (\ref{ladder_diagram}) is decoupled. We can assume the form of the corrected velocity $\widetilde{v}^{\alpha}_{y\bold{k}}=\eta v_{y\bold{k}}^{\alpha}$ ($\alpha=K,-K$) and obtain that
\begin{eqnarray}
\eta&=&\frac{1}{1-w_1^2\chi_1^2(\tau/\tau_{(1,1,1)})}.
\end{eqnarray}
We can see that the intervalley scattering enters $\eta$ only through $\tau$, which is defined in Eq.~(\ref{tau}). $\eta$ gets suppressed ($\rightarrow 1$) by the intervalley scattering as $1/\tau_{(1,2,2)}\gg 1/\tau_{(1,1,1)}$. The same calculation applies to the spin current operator $j_x^z$, and we have the corrected $\widetilde{j}_x^z=\eta j_x^z$ at each valley. Numerical results are shown in Fig. \ref{fig:vertex1}, where $\eta$ increases from 1 as the Fermi energy moves away from the valence band edge. Different from usual multi-band systems, the correction here does not modify the intrinsic SHC directly, and the reason is due to the conservation of valley index mentioned above. Later we will show that the correction is manifested through the extrinsic spin Hall effect.

\begin{figure}[htbp]
\centering
\includegraphics[width=\columnwidth]{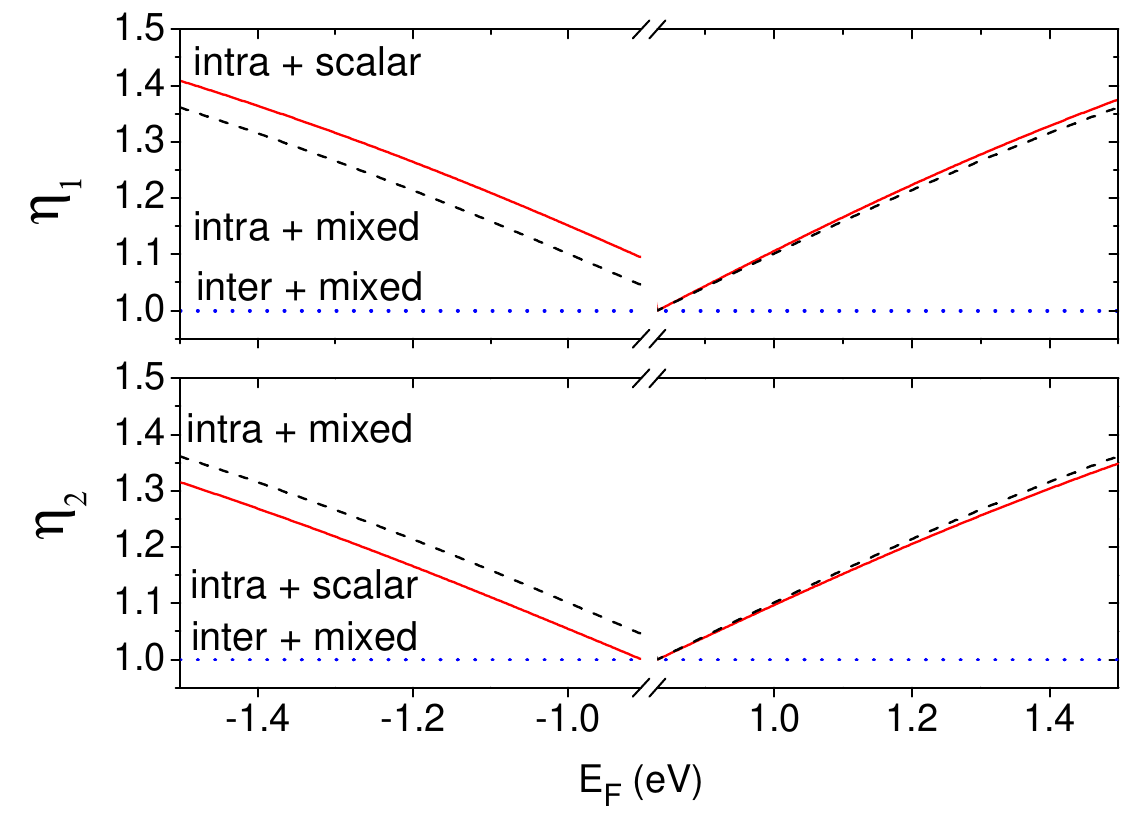}
\caption{Vertex correction factors $\eta_1$ and $\eta_2$ as functions of the Fermi energy $E_F$, in the presence of the scalar potential induced intravalley scattering (solid), intravalley scattering with the equal contribution from the scalar and magnetic potentials (dashed), and intervalley scattering with the equal scalar and magnetic contributions (dotted). All bands in both $K$ and -$K$ valleys are included. The positive and negative $E_F$ correspond to electron- and hole-doped regimes, respectively. All parameters are adopted for MoS$_2$ from Ref. [\onlinecite{Xiao12prl}].}\label{fig:eta12}
\end{figure}

\emph{Multi-band case}: We now consider the multiband effect on vertex correction when the system becomes electron- or heavily hole-doped. The forms of relaxation times in this case are shown in Appendix \ref{sec:relaxation1}. We assume that $\widetilde{v}^{n}_{y\bold{k}}=\eta_n v_{y\bold{k}}^{n}$ ($n=1,2$ for $\tau s=\pm1$).  It can be demonstrated that
\begin{eqnarray}
\left(\begin{array}{cc}
\eta_1 \\
\eta_2 \\
\end{array}\right)&=&\left(\begin{array}{cc}
1 \\
1 \\
\end{array}\right)+
\left(\begin{array}{cc}
\tau_{K,\up}/\tau_{(1,1,1)} & \tau_{K,\dn}/\tau_{(2,2,1)} \\
\tau_{K,\up}/\tau_{(1,2,1)} & \tau_{K,\dn}/\tau_{(2,1,1)} \\
\end{array}\right)\nonumber\\
&\times&\left(\begin{array}{cc}
\chi_1^2w_1^2\eta_1 \\
\chi_2^2w_2^2\eta_2 \\
\end{array}\right).
\end{eqnarray}
After solving these equations, $\eta_{1,2}$ can be derived. The same argument can be applied to the spin current operator $\widetilde{j}_{x\bold{k}}^{z,n}=\eta_nj_{x\bold{k}}^{z,n}$. The numerical results for $\eta_{1}$ and $\eta_2$ are given in Fig. \ref{fig:eta12}, and three different cases are compared: pure intravalley scalar scattering, pure intravalley (intervalley) scattering with equal scalar and magnetic contributions.
For the intravalley scattering, $\eta_{1}$ and $\eta_2$ increase as the Fermi energy moves away from the band edges, much like in the lightly hole-doped regime. In contrast, there is no correction ($\eta_{1,2}=1$) for the pure intervalley scattering due to the loss of coherence.

\section{\label{sec:spin_Hall_hole}Spin Hall conductivity of the valence band at valley K}

The SHC $\sigma_{xy}^z$ is the response function of spin current
\begin{eqnarray}
j_x^z&=&\frac{\hbar}{4}\{v_x,\sigma_z\otimes I\}
\end{eqnarray}
to the charge current $j_y$. Note that in spin-orbit coupled systems, the spin in general is not a conversed quantity.~\cite{Shi06prl}  However, in $MX_2$ monolayers, because of the in-plane mirror symmetry,  the $z$-component of the spin is conserved and the above definition is valid.  Similar to the anomalous Hall effect,\cite{Sinitsyn07prb,Yang11prb,Nagaosa10rmp} the spin Hall effect has both intrinsic and extrinsic contributions. The SHC can be derived from the Kubo-Streda formula.\cite{Crepieux01prb,Streda82jpc} In the weak scattering limit, the conductivity is separated into
$\sigma_{xy}^z=\sigma_{xy}^{z,\mathrm{I}}+\sigma_{xy}^{z,\mathrm{II}}$, where $\sigma_{xy}^{z,\mathrm{I}}$, $\sigma_{xy}^{z,\mathrm{II}}$ are the contribution near the Fermi surface and intrinsic contribution from the Fermi sea, respectively. The intrinsic SHC is independent of disorder, but determined by the Berry curvature of occupied states (the Fermi sea contribution).\cite{Sundaram99prb,Xiao10rmp} On the other hand, the extrinsic SHC is given by the disorder-dependent part of $\sigma_{xy}^{z,\mathrm{I}}$ term
\begin{eqnarray}
\sigma_{xy}^{z,I}&=&\frac{e\hbar}{2\pi S}\mathrm{Tr}\langle j_x^zG^R(E_F)v_yG^A(E_F)\rangle,
\end{eqnarray}
where $G^{R/A}$ is the retarded (advanced) Green's function dressed by the impurity scattering. $e$ is the electron charge. For convenience, we will multiply a factor $2e/\hbar$ to the SHC, so that it has the units of the charge conductivity. In this section, we focus on the lightly hole-doped regime as shown in Fig. \ref{fig:band_dispersion}, when the Fermi energy intersects just a single band at each valley.

\subsection{\label{sec:intrinsic}Intrinsic spin Hall conductivity}

In monolayer $MX_2$, the $z$-component of the spin is conserved in each band, so the derivation of the SHC is equivalent to two copies of the anomalous Hall conductivity. The intrinsic anomalous Hall effect originates from the Berry curvature\cite{Xiao10rmp}
\begin{eqnarray}
\Omega_n^z(\bold{k})&=&\hat{z}\cdot\nabla_{\bold{k}}\times  \langle u_n(\bold{k})|i\nabla_{\bold{k}}|u_n(\bold{k})\rangle
\end{eqnarray}
of occupied states, where $u_n(\bold{k})$ is the eigenfunction for band $n$ and wave vector $\mathbf{k}$, and $\hat{z}$ the unit vector along the $z$ axis. Time-reversal symmetry requires that $\Omega_n(-\bold{k})=-\Omega_n(\bold{k})$, leading to opposite anomalous Hall conductivity for different valleys. When combined with the spin-valley coupled property, each valley gives the same contribution to the SHC. The intrinsic spin Hall effect has been studied by first-principles calculations for this system.\cite{Feng12prb}

At zero temperature, the intrinsic SHC is given by
\begin{eqnarray}
\sigma^{int}_{xy}=\frac{e^2}{h}\sum_{n=1,2} (-1)^n \int \frac{d^2\mathbf{k}}{(2\pi)^2}\Omega^z_n(\mathbf{k})\Theta(E_F - E_\mathbf{k}),
\end{eqnarray}
where at valley $K$, $n=1,2$ corresponds to spin up and down, respectively.
Based on the low-energy effective model, the intrinsic SHC is found as
\begin{eqnarray}
\sigma^{int}_{xy}&=&\frac{e^2}{2h}(1-\cos\theta_{1}),
\end{eqnarray}
where $\theta_1$ is defined by Eq. (\ref{theta_def}) on the Fermi surface. Obviously, $\sigma^{int}_{xy}$ reaches its maximum at the band edge of the second highest valence band.

\subsection{\label{sec:extrinsic}Extrinsic spin Hall conductivity}

\begin{figure}[htbp]
\centering
\includegraphics[width=0.8\columnwidth]{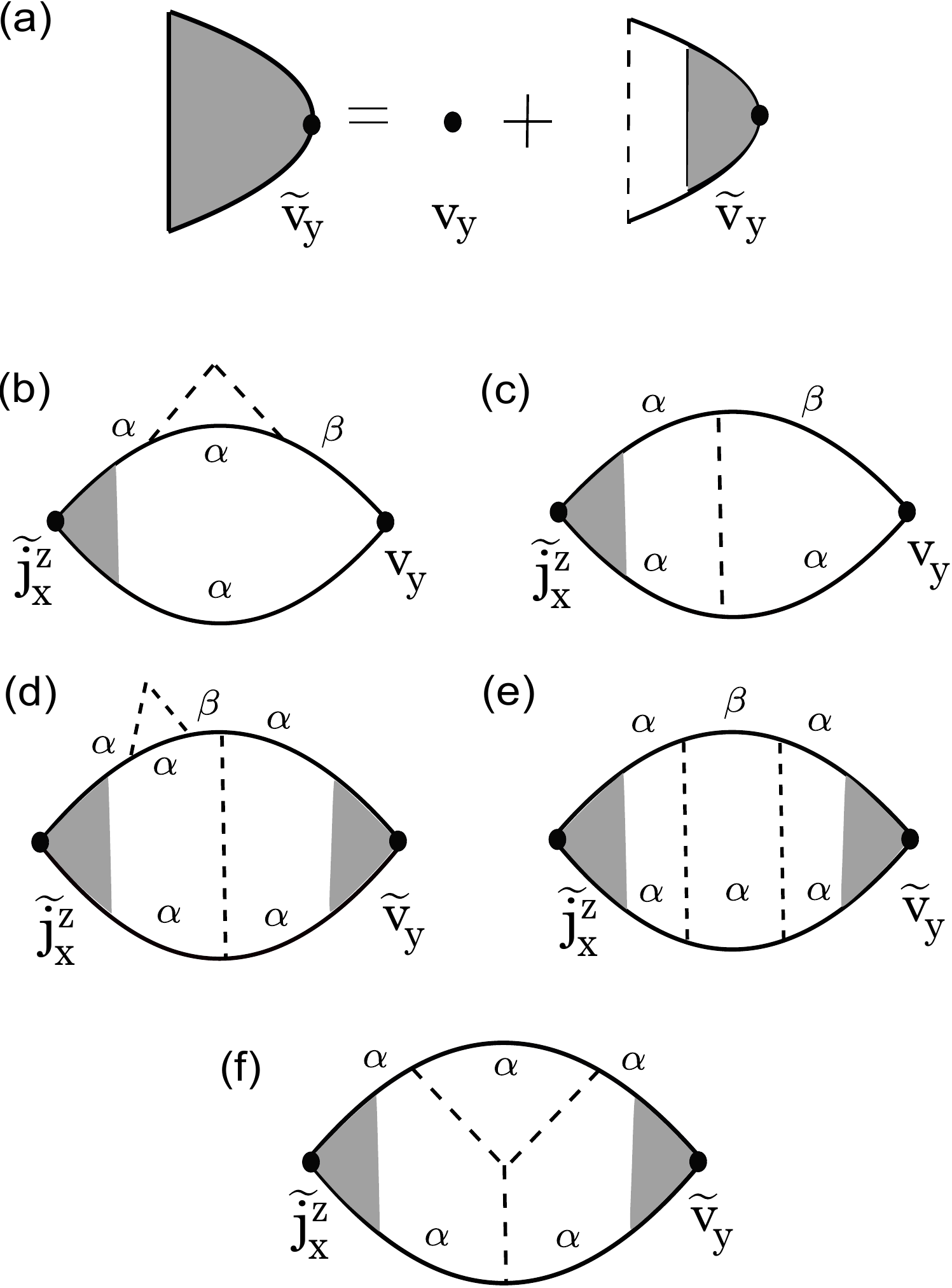}
\caption{ (a) Ladder-diagram correction to velocity vertex shown in grey region. (b)-(f) Diagrams contributing to the extrinsic spin Hall conductivity. $\alpha$ and $\beta$ denote different bands and dashed lines refer to correlated disorder scattering.}\label{fig:diagram}
\end{figure}

The extrinsic spin Hall effect comes from electrons near the Fermi surface when they are scattered by impurities and disorders, and can be divided into from side-jump \cite{Berger70prb} and skew-scattering mechanisms.\cite{Smit55physica} The Feynman diagrams to calculate them are depicted in Fig. \ref{fig:diagram} (b)-(f).\cite{Sinitsyn07prb,Yang11prb,Lu13arXiv} In the semiclassical picture, side-jump terms can be further classified into three contributions. (1) From the anomalous distribution function in Fig. \ref{fig:diagram} (b) and (c); (2) From the coordinate shift by making a $180^\circ$ rotation of Fig. \ref{fig:diagram} (b) and (c) followed by exchanging symbols $j^z_x$ and $v_y$; (3) skew scattering-induced side jump in Fig. \ref{fig:diagram} (d) and (e). The third-order correlation-related skew scattering is drawn in Fig. \ref{fig:diagram} (f). For each diagram in Fig. \ref{fig:diagram} (b)-(f), there exists a symmetric copy by exchanging $\alpha$ and $\beta$. \cite{Sinitsyn07prb,Yang11prb} Here $\alpha$ and $\beta$ denotes different combinations of indices including valley, spin, and band.

Diagrams in Fig.~\ref{fig:diagram} (b)-(e) all contain at least one asymmetric correlation function $\langle U^{\alpha\alpha}U^{\alpha\beta}\rangle_{dis}$, which is angle-dependent and may lead to nonvanishing results. However, for the valley index, the correlations $\langle U^{KK}U^{K-K}\rangle_{dis}$,$\langle U^{-K-K}U^{-KK}\rangle_{dis}$ are forbidden due to the violation of the valley conservation. This means that different valleys are decoupled in the side-jump mechanism and can be treated separately.

Back to the spin index, the correlation functions $\langle U^{\up\up}U^{\up\dn}\rangle_{dis}$, $\langle U^{\dn\dn}U^{\dn\up}\rangle_{dis}$ are neglected since the scattering in different directions is assumed to be uncorrelated. This implies that we can further decouple the spin part in the side-jump mechanism even when the scattering is spin-dependent. Hence we can limit $\alpha$ and $\beta$ to only band index.  This is supported by the fact that the interband scattering can contribute to the spin Hall effect via virtual interband transitions. Now the calculation becomes similar to that for the anomalous Hall effect,\cite{Sinitsyn07prb} and side-jump Hall conductivity at each valley has the form
\begin{equation} \begin{split}
\sigma^{sj}_{xy}&=-\frac{e^2}{2h}\eta\sin^2\theta_1\cos\theta_1\frac{\tau}{\tau_{(1,1,1)}} \\
&\quad \times (1+\frac{3}{16}\eta\sin^2\theta_1\frac{\tau}{\tau_{(1,1,1)}}),
\end{split} \end{equation}
while the skew-scattering Hall conductivity reads
\begin{eqnarray}
\sigma^{sk}_{xy}&=&\frac{e^2}{8h}\eta^2\sin^4\theta_1\cos\theta_1
(\frac{\tau}{\tau_{(1,1,3)}})^2.
\end{eqnarray}
It is clear that $\sigma^{sj}_{xy}$ has opposite sign compared with $\sigma^{int}_{xy}$, while $\sigma^{sk}_{xy}$ shows the same sign, which means that the skew scattering enhances the intrinsic spin-Hall effect while the side jump suppresses it. Note that $\sigma^{int}_{xy}$ and $\sigma^{sj}_{xy}$ are independent of the total disorder concentration, while $\sigma^{sk}_{xy}\sim n^{-1}$ following the definition of $\tau$ in Eq.~\eqref{tau} and $\tau_{(1,1,3)}$ in Eq.~\eqref{a9}. This implies that in the clean limit the skew scattering becomes dominant. On the other hand, the scalar and magnetic scattering do not make much difference since the spin part is decoupled in the side-jump mechanism. The only difference is that magnetic scattering can contribute to the intervalley scattering, and thus modify the total scattering time.

\subsection{\label{sec:total1}Total contribution}

\begin{figure}[htbp]
\centering
\includegraphics[width=\columnwidth]{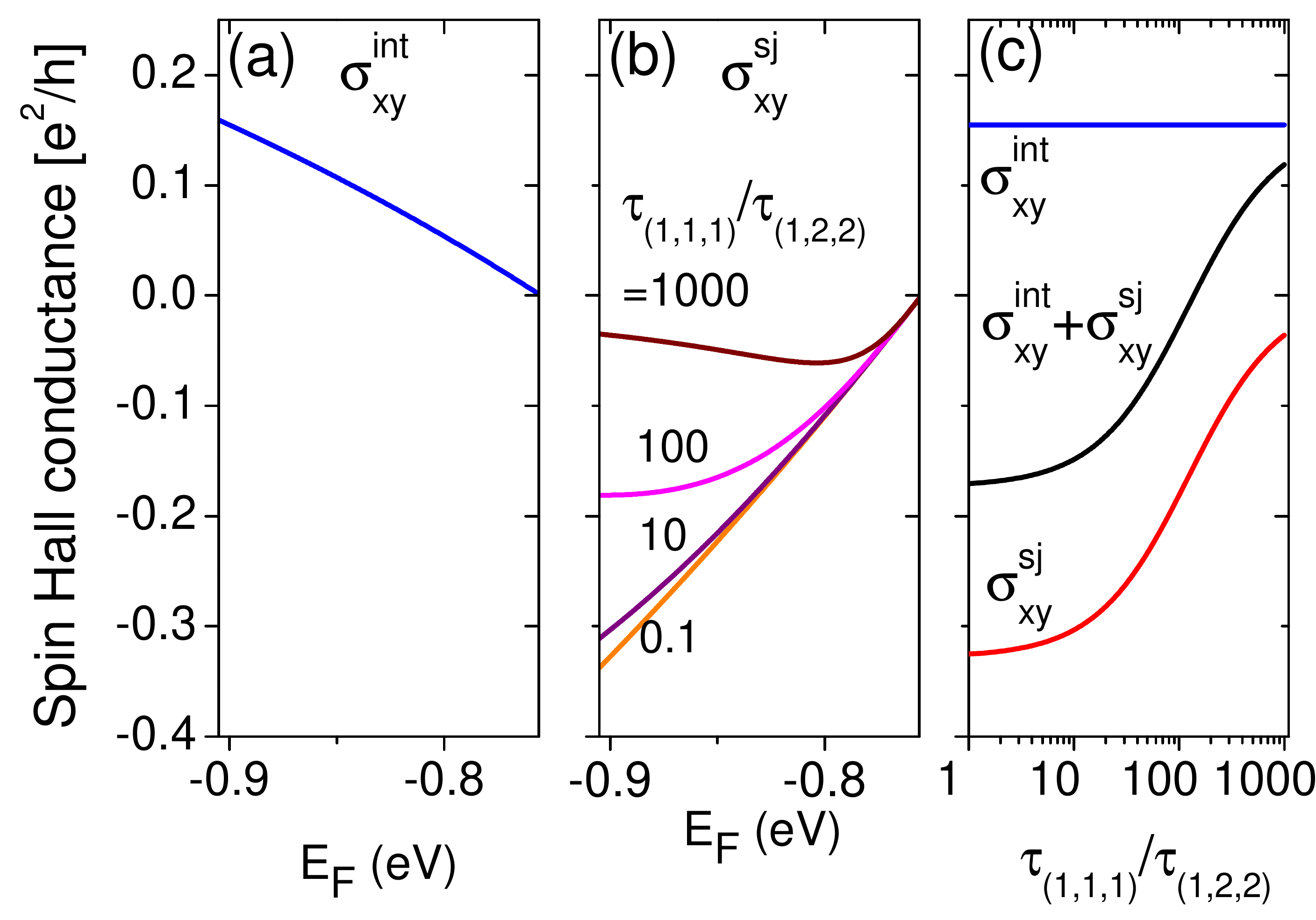}
\caption{The spin Hall conductivity in the lightly hole-doped regime. (a) The intrinsic spin Hall conductivity as a function of the Fermi energy $E_F$.
(b) The side-jump spin Hall conductivity vs $E_F$ for different $\tau_{(1,1,1)}/\tau_{(1,2,2)}$, the ratio of the intervalley scattering to intravalley scattering. (c) The intrinsic and side-jump spin Hall conductivities as functions of $\tau_{(1,1,1)}/\tau_{(1,2,2)}$ at $E_F=-0.90eV$. All parameters are adopted for MoS$_2$ from Ref. [\onlinecite{Xiao12prl}]. }\label{fig:result1}
\end{figure}

The total SHC in the lightly hole-doped monolayer $MX_2$ reads
\begin{eqnarray}\label{total}
\sigma^z_{xy}&=&2\times(\sigma^{int}_{xy}+\sigma^{sj}_{xy}+\sigma^{sk}_{xy}),
\end{eqnarray}
where the factor $2$ comes from the valley degeneracy. At a low doping level when $|E_F|\ll\Delta$, $\sigma^{int}_{xy}$ and $\sigma^{sj}_{xy}$ are of the order of $O(\Delta^{-2})$, while $\sigma^{sk}_{xy}$ is of the order of $O(\Delta^{-3},n^{-1})$. This means that $\sigma^{sk}_{xy}$ only dominates in the ultraclean limit [$\sigma_{xx}>10^6(\Omega$ cm$)^{-1}$],\cite{Nagaosa10rmp} otherwise this term can be safely neglected. For the realistic parameters of MoS$_2$\cite{Xiao12prl} with hole-doped carrier density $n_h=1.0\times10^{13}$cm$^{-2}$ and mobility $\mu=200$ cm$^{2}$V$^{-1}$s$^{-1}$,\cite{Radisavljevic11naturenano} we can write down a three-dimensional version of longitudinal conductivity $\sigma_{xx}a^{-1}$ for comparison, where $a=3.193\AA$ is the lattice constant. This gives $\sigma_{xx}a^{-1}=en_h\mu a^{-1}=10^4(\Omega$ cm$)^{-1}$, which implies that skew scattering can be neglected. The results for the intrinsic and side-jump contributions are shown in Fig. \ref{fig:result1}.  It is clear that these two terms always have opposite signs. Both the intrinsic ($\sigma_{xy}^{int}$) and side-jump ($\sigma_{xy}^{sj}$) contributions depend on the Fermi energy [Figs. \ref{fig:result1} (a) and \ref{fig:result1} (b)], while the side-jump contribution also depends on the ratio [Fig. \ref{fig:result1} (b)]
\begin{eqnarray}
\frac{\tau_{(1,1,1)}}{\tau_{(1,2,2)}} = \frac{n_m t_x^2+n_mt_y^2}{n_0u_0^2+n_m u_z^2},
\end{eqnarray} which measures the energy-independent scattering ratio between the inter- and intravalley scattering. $\sigma^{sj}_{xy}$ can be suppressed by the intervalley scattering. As a result, by tuning $\tau_{(1,1,1)}/\tau_{(1,2,2)}$, the total SHC, as a summation of the intrinsic and side-jump contributions, could change sign. This may offer a new way to estimate the strength of the intervalley scattering in $MX_2$ monolayers.

\section{\label{sec:spin_Hall_electron} Multi-band Spin Hall conductivity}

In this section we extend the above discussion to the multiband case, where at each valley, two bands contribute to the SHC. This situation corresponds to the electron- or heavily hole-doped cases.  The vertex correction in this regime has been discussed in Sec.~\ref{sec:disorder_relaxation_vertex}.

\subsection{\label{sec:intrinsic2}Intrinsic spin Hall conductivity}

In the electron-doped regime, the intrinsic SHC reads
\begin{eqnarray}
\sigma^{int}_{xy}&=&\frac{e^2}{2h}(\cos\theta_1-\cos\theta_2),
\end{eqnarray}
where $\theta_{1,2}$ are defined in Eq. (\ref{theta_def}) on the Fermi surface. When tuned to the heavily hole-doped regime, the SHC becomes
\begin{eqnarray}
\sigma^{int}_{xy}&=&-\frac{e^2}{2h}(\cos\theta_1-\cos\theta_2).
\end{eqnarray}

\subsection{\label{sec:extrinsic2}Extrinsic spin Hall conductivity}

In the diagrammatic language, the side-jump contribution $\sigma^{sj}_{xy}$ comes from the asymmetric scattering correlation.  However, in the present case the two bands at each valley have opposite spin-polarization and thus the asymmetric correlation between them must vanish. As a result, $\sigma^{sj}_{xy}$ is contributed independently by each band, which reads
\begin{equation} \begin{split}
\sigma^{sj}_{xy}&=-\frac{e^2}{2h}[\eta_1\sin^2\theta_1\cos\theta_1\frac{\tau_{\up,K}}{\tau_{(1,1,1)}}\\
&\quad\times(1+\frac{3\eta_1}{16}\sin^2\theta_1\frac{\tau_{\up,K}}{\tau_{(1,1,1)}}) \\
&\quad-\eta_2\sin^2\theta_2\cos\theta_2\frac{\tau_{\dn,K}}{\tau_{(2,1,1)}}\\
&\quad\times(1+\frac{3\eta_2}{16}\sin^2\theta_2\frac{\tau_{\dn,K}}{\tau_{(2,1,1)}})],
\end{split} \end{equation}
for both electron and heavily hole-doped regime. Since there are more channels now, the relaxation times become band-dependent and the expressions can be found in Appendix \ref{sec:relaxation1}. In general, time-reversal symmetry requires that $\tau_{\up,-K}=\tau_{\dn,K}$ and $\tau_{\dn,-K}=\tau_{\up,K}$.

As for the skew scattering, the argument is essentially the same and we have
\begin{eqnarray}
\sigma^{sk}_{xy}&=&\mp\frac{e^2}{8h}[\eta_1^2\sin^4\theta_1\cos\theta_1
(\frac{\tau_{\up,K}}{\tau_{(1,1,3)}})^2\nonumber\\
&-&\eta_2^2\sin^4\theta_2\cos\theta_2
(\frac{\tau_{\dn,K}}{\tau_{(2,1,3)}})^2],
\end{eqnarray}
where $\mp$ refers to electron- and heavily hole-doped regime, respectively.

\subsection{\label{sec:total2}Total contribution}

\begin{figure}[htbp]
\centering
\includegraphics[width=\columnwidth]{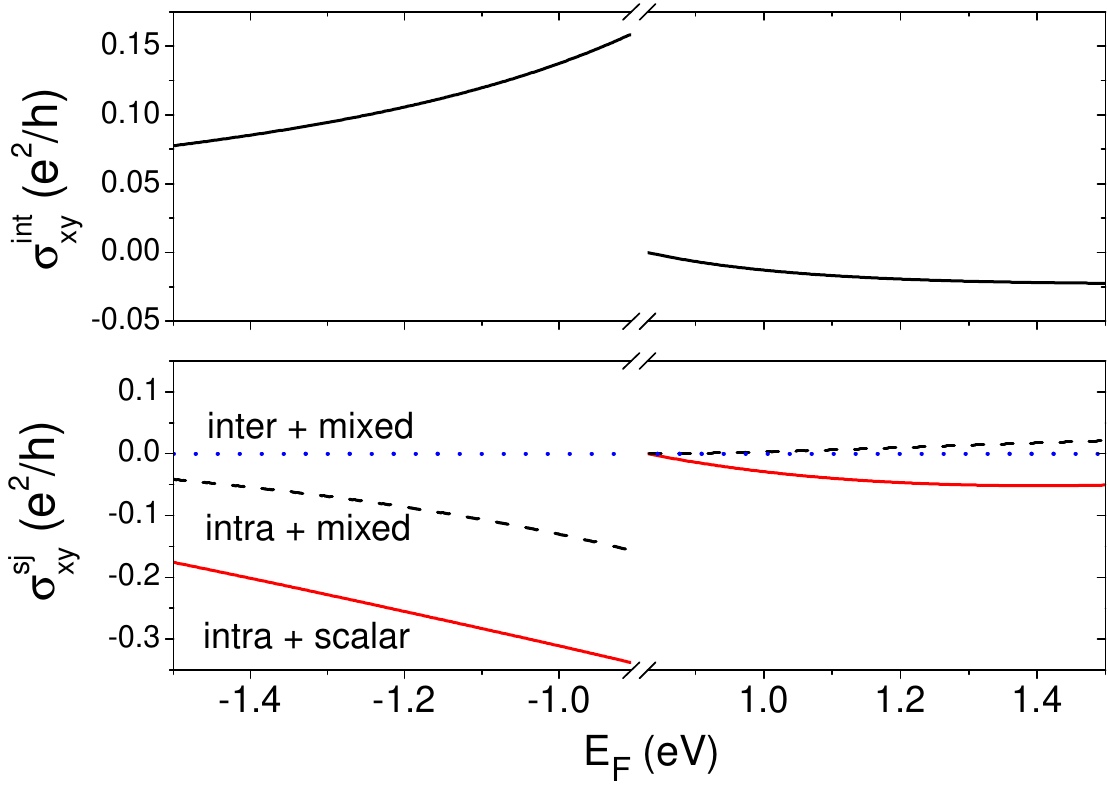}
\caption{The intrinsic ($\sigma_{xy}^{int}$) and side-jump ($\sigma_{xy}^{sj}$) spin Hall conductivity in the electron- and heavily hole-doped regime as functions of the Fermi energy $E_F$. Solid, dashed, and dotted lines correspond to the cases with the scalar potential induced intravalley scattering, intravalley scattering with the equal contribution from the scalar and magnetic potentials, and intervalley scattering with the equal scalar and magnetic contributions, respectively. All parameters are adopted for MoS$_2$ from Ref. [\onlinecite{Xiao12prl}].}\label{fig:sigma-multi-EF}
\end{figure}

\begin{figure}[htbp]
\centering
\includegraphics[width=\columnwidth]{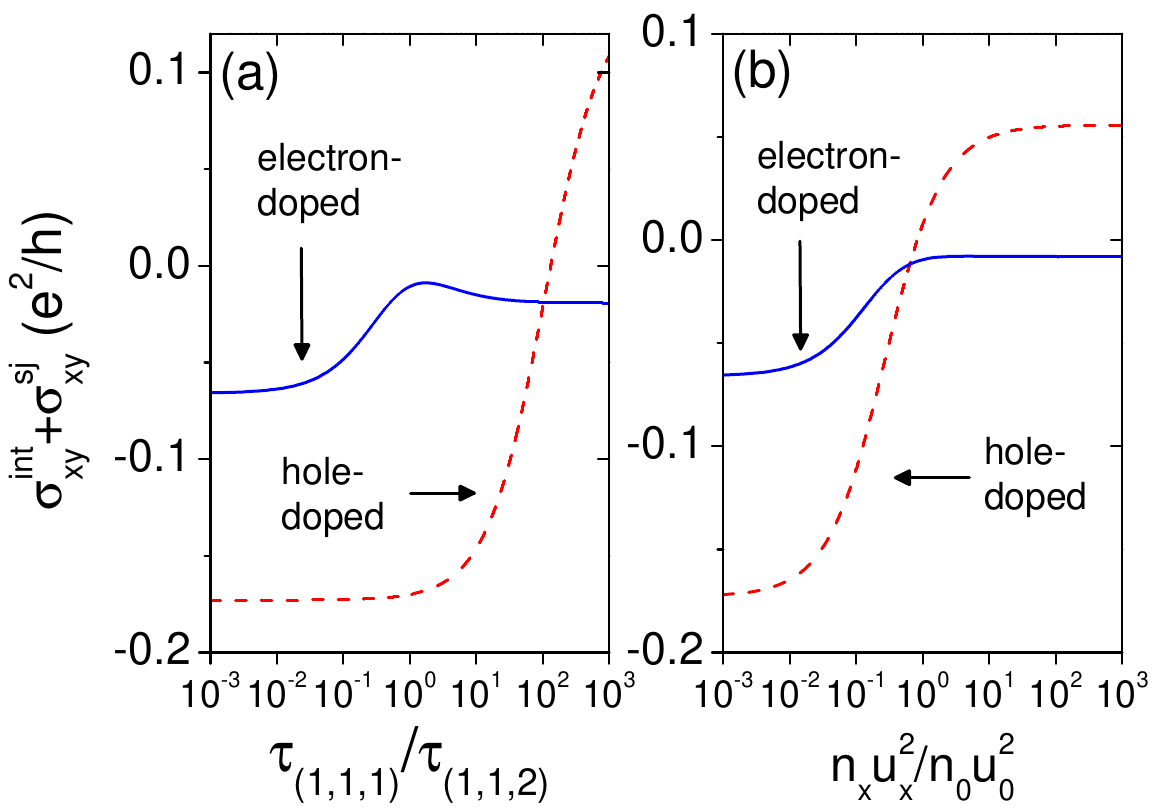}
\caption{The total spin Hall conductivity $\sigma_{xy}^{int}+\sigma_{xy}^{sj}$ as a function of (a) ratio of intra- to intervalley scattering time
and (b) ratio of scalar to magnetic scattering. $E_F=1.2$ eV for electron-doped case (solid), and $E_F=-1.0$ eV for hole-doped case (dashed). In (a), the magnetic scattering is absent, and in (b) the intervalley scattering is absent. All parameters are adopted for MoS$_2$ from Ref. [\onlinecite{Xiao12prl}].}\label{fig:sigma-tau-n}
\end{figure}

The total contribution is given by the summation of $\sigma^{int}_{xy}$ and $\sigma^{sj}_{xy}$; $\sigma^{sk}_{xy}$ is neglected. Three different cases are compared in Fig. \ref{fig:sigma-multi-EF}: pure intravalley scalar scattering, pure intravalley (intervalley) scattering with equal scalar and magnetic contributions. Obviously, the SHC has a much larger value in hole-doped regime than in electron-doped regime, due to the existence of large spin splitting in the valence band. Similar to the conclusion in the last section, by tuning the ratio of intra- to intervalley scattering time, i.e. $\tau_{(1,1,1)}/\tau_{(1,1,2)}$, the SHC exhibits a sign change in the hole-doped regime in Fig. \ref{fig:sigma-tau-n} (a), while this is not the case for the electron-doped regime. Moreover, different from the single-band case where the intervalley and magnetic scattering are locked, here the intervalley and magnetic scattering can be tuned independently. As a result, it is found that by tuning the ratio of the scalar to magnetic scattering time, i.e., $n_xu_x^2/n_0u_0^2$, the SHC again exhibits a sign change in the hole-doped regime, as shown in Fig. \ref{fig:sigma-tau-n} (b).

\section{\label{sec:discussion1}Discussion and conclusion}

As discussed in Secs.~\ref{sec:spin_Hall_hole} and \ref{sec:spin_Hall_electron}, the side-jump and intrinsic contributions are comparable with each other, highlighting the important role of disorder in the spin Hall effect in monolayer $MX_2$. Consider a hole-doped MoS$_2$ sample with a carrier density of $n_h=1.0\times10^{13}$cm$^{-2}$, for which the Fermi energy lies in the spin-split valence bands.  In the absence of the intervalley scattering, the intrinsic and side-jump contributions are $\sigma_{int}=0.90\times10^{-2}e^2/\hbar$ and $\sigma_{sj}=-1.83\times10^{-2}e^2/\hbar$, respectively. As a result, the total SHC becomes $\sigma_{xy}^z=-0.93\times10^{-2}e^2/\hbar$, which is comparable with those in semiconductors GaAs, Ge and AlAs.\cite{Yao05prl,Guo05prl} Experimentally, the SHC can be evaluated by fitting the measured spin accumulation at edges.

Although our calculations are mainly based on the low-energy effective model at $\pm K$, the conclusion is valid even when the $\Gamma$ valley\cite{Cheiwchanchamnangij12prb,Shi12arXiv,Ellis11apl} is involved. The reason is that the existence of the large effective mass and small spin splitting at $\Gamma$ valley results in a negligible spin-Hall conductivity.

In summary, we have studied the spin Hall conductivity of monolayer MoS$_2$ with both intrinsic and extrinsic contributions. We find that in this large-gap system the side-jump contribution is comparable with the intrinsic contribution. The side-jump and intrinsic contributions have opposite signs. The side-jump contribution can be suppressed by the intervalley scattering. By tuning the ratio of intra- to intervalley scattering, the total spin Hall conductivity shows a sign change in hole-doped samples, which can be used to measure the strength of the intervalley scattering.

We acknowledge useful discussion with Dimitrie Culcer.  This work was supported by the U.S. Department of Energy, Office of Basic Energy Sciences, Materials Sciences and Engineering Division (W.S.) and by AFOSR Grant No.~FA9550-12-1-0479 (D.X.)

\appendix

\section{\label{sec:relaxation1}Relaxation time}

We define a set of relaxation times
\begin{eqnarray}
\frac{1}{\tau_{(n,1,1)}}&=&\frac{2\pi}{\hbar}N_n(n_0u_0^2+n_zu_z^2),\\
\frac{1}{\tau_{(n,2,1)}}&=&\frac{2\pi}{\hbar}N_n(n_xu_x^2+n_yu_y^2),\\
\frac{1}{\tau_{(n,1,2)}}&=&\frac{2\pi}{\hbar}N_n(n_0t_0^2+n_zt_z^2),\\
\frac{1}{\tau_{(n,2,2)}}&=&\frac{2\pi}{\hbar}N_n(n_xt_x^2+n_yt_y^2),\   \ n=1,2
\end{eqnarray}
which are functions of scattering potential, disorder concentration and density of states. Then the relaxation time in multiband cases can be conveniently expressed by using these new definitions. For example, in the electron-doped regime at valley $K$ the relaxation time reads
\begin{eqnarray}
\frac{1}{\tau_{\up,K}}&=&\frac{1}{\tau_{(1,1,1)}}(\chi_1^4+w_1^4)+\frac{1}{\tau_{(2,2,1)}}(\chi_1^2\chi_2^2+w_1^2w_2^2)\nonumber\\
&+&\frac{1}{\tau_{(2,1,2)}}\chi_1^2\chi_2^2+\frac{1}{\tau_{(1,2,2)}}\chi_1^4,\\
\frac{1}{\tau_{\dn,K}}&=&\frac{1}{\tau_{(2,1,1)}}(\chi_2^4+w_2^4)+\frac{1}{\tau_{(1,2,1)}}(\chi_1^2\chi_2^2+w_1^2w_2^2)\nonumber\\
&+&\frac{1}{\tau_{(1,1,2)}}\chi_1^2\chi_2^2+\frac{1}{\tau_{(2,2,2)}}\chi_2^4.
\end{eqnarray}
And time-reversal symmetry guarantees that $\tau_{\up,-K}=\tau_{\dn,K}$ and $\tau_{\dn,-K}=\tau_{\up,K}$. Similarly in the heavily hole-doped regime we have
\begin{eqnarray}
\frac{1}{\tau_{\up,K}}=\frac{1}{\tau_{\dn,-K}}&=&\frac{1}{\tau_{(1,1,1)}}(\chi_1^4+w_1^4)
+\frac{1}{\tau_{(2,1,2)}}w_1^2w_2^2\nonumber\\
&+&\frac{1}{\tau_{(2,2,1)}}(\chi_1^2\chi_2^2+w_1^2w_2^2)+\frac{1}{\tau_{(1,2,2)}}w_1^4,\nonumber\\
\\
\frac{1}{\tau_{\dn,K}}=\frac{1}{\tau_{\up,-K}}&=&\frac{1}{\tau_{(2,1,1)}}(\chi_2^4+w_2^4)
+\frac{1}{\tau_{(1,1,2)}}w_1^2w_2^2\nonumber\\
&+&\frac{1}{\tau_{(1,2,1)}}(\chi_1^2\chi_2^2+w_1^2w_2^2)+\frac{1}{\tau_{(2,2,2)}}w_2^4.\nonumber\\
\end{eqnarray}

Also, the relaxation times for the skew scattering can be given as \cite{Yang11prb}
\begin{eqnarray}
\frac{1}{\tau^2_{(1,1,3)}}&=&\frac{4\pi^3a^2t^2}{\hbar^2}N_1^3(n_0v_0^3+n_zv_z^3), \label{a9}\\
\frac{1}{\tau^2_{(2,1,3)}}&=&\frac{4\pi^3a^2t^2}{\hbar^2}N_2^3(n_0v_0^3+n_zv_z^3),
\end{eqnarray}
and it is clear that $1/\tau^2_{(1,1,3)}$ and $1/\tau^2_{(2,1,3)}$ are proportional to disorder concentration $n_{dis}$.

\end{document}